\documentclass[pra,twocolumn,showpacs, superscriptaddress,10pt]{revtex4-1}
\usepackage{amsmath,amssymb,amsfonts}
\usepackage{graphicx} 
\usepackage{txfonts}
\usepackage{color}
\usepackage{bm}
\usepackage{enumerate}

\usepackage[unicode,breaklinks]{hyperref}
\hypersetup{
    unicode=true,
    a4paper=true,
    plainpages=false,
    pdftitle={Persistent current formation in a high-temperature Bose-Einstein condensate},
    pdfauthor={S. J. Rooney, T. W. Neely, B. P. Anderson, and A. S. Bradley},
    pdfsubject={Persistent current formation in a high-temperature Bose-Einstein condensate},
    colorlinks=true,
    linkcolor=blue,
    citecolor=blue,
    filecolor=black,
    urlcolor=blue
}
\newcommand{\eref}[1]{(\ref{#1})} 
\newcommand{\eeref}[1]{Eq.~(\ref{#1})} 
\newcommand{\fref}[1]{Fig.~\ref{#1}}
\newcommand{\x}{\mathbf{r}}
\newcommand{\eqn}[1]{\begin{eqnarray}#1 \end{eqnarray}}
\newcommand{\ecut}{\epsilon_{\rm{cut}}}
\newcommand{\of}[1]{(#1)}
\newcommand{\rC}{\textbf{C}} 
\newcommand{\rI}{\textbf{I}} 
\newcommand{\PC}{\mathcal{P}}

\begin{document}

\title{Persistent current formation in a high-temperature Bose-Einstein condensate:\\ an experimental test for c-field theory}
\author{S. J. Rooney} 
\affiliation{Jack Dodd Center for Quantum Technology, Department of Physics, University of Otago, Dunedin 9016, New Zealand.}
\author{T. W. Neely}
\altaffiliation[Current address: ]{School of Mathematics and Physics, University of Queensland, QLD 4072, Australia.}
\affiliation{College of Optical Sciences, University of Arizona, Tucson, Arizona 85721, USA.}

\author{B. P. Anderson} 
\affiliation{College of Optical Sciences, University of Arizona, Tucson, Arizona 85721, USA.}
\author{A.~S. Bradley} 
\affiliation{Jack Dodd Center for Quantum Technology, Department of Physics, University of Otago, Dunedin 9016, New Zealand.}
\date{\today}


\begin{abstract}
Experimental stirring of a toroidally trapped Bose-Einstein condensate at high temperature generates a disordered array of quantum vortices that decays via thermal dissipation to form a macroscopic persistent current [T. W. Neely {\em et al.} \href{http://arxiv.org/abs/1204.1102}{arXiv:1204.1102} (2012)].  We perform 3D numerical simulations of the experimental sequence within the Stochastic Projected Gross-Pitaevskii equation using \emph{ab initio} determined reservoir parameters. We find that both damping \emph{and} noise are essential for describing the dynamics of the high-temperature Bose field. The theory gives a quantitative account of the formation of a persistent current, with no fitted parameters.
\end{abstract}
\maketitle
\section{Introduction}
Since the first observations of quantized vortices in dilute gas Bose-Einstein condensates (BECs)~\cite{Matthews1999}, experimental studies of vortices have proliferated~\cite{Anderson:2010bd}. Vortex motion is highly sensitive to thermal fluctuations and nonlinear interactions between vortices and other excitations in the fluid~\cite{Pismen1999}, and many experiments depend on the presence~\cite{Haljan2001} or formation~\cite{Madison2000} of a rotating thermal reservoir to take the superfluid from a non-rotating state to one containing vortices. In an annular trapping geometry, such as may be created using an obstacle potential in a harmonic trap~\cite{Weiler:2008eu}, a vortex can be pinned to the obstacle to create a BEC in a state of perpetual motion, forming a \emph{persistent current}~\cite{Ryu07a,Weiler:2008eu,Ramanathan:2011bi}. In the absence of coherent optical manipulation~\cite{Ryu07a,Ramanathan:2011bi}, the dynamical evolution from a non-rotating ground state to one containing a topologically stabilized superflow necessitates the motion of vortices toward the inner boundary of the toroidal system, through the dual action of \emph{forcing} and \emph{dissipation}, requiring a thermal reservoir of non-condensed atoms to drive this process. 

In this article we present a study of persistent current formation via dissipative vortex dynamics in the presence of a large, high-temperature thermal reservoir. While dissipative vortex dynamics in harmonically trapped BECs have been treated numerically~\cite{Jackson:2009jo,Allen:2013cs,Weiler:2008eu,Rooney:2010dp,Rooney:2011fm}, the predictions for vortex lifetimes have not yet been tested experimentally. Here we perform large-scale numerical simulations of an experimental forcing sequence that generates a long-lived persistent current via forcing and thermal dissipation~\cite{Neely:2012vj}. To describe the conditions of the experiment we use the Stochastic Projected Gross-Pitaevskii equation (SPGPE)~\cite{Gardiner:2003bk,Bradley:2008gq,Rooney:2012gb} for the dynamical evolution of a system of partially coherent matter waves in contact with a thermal reservoir of high-energy, incoherent atoms. We  compare the SPGPE and damped GPE (dGPE)~\cite{Choi:1998eh,Penckwitt2002,Tsubota2002} (obtained by neglecting the noise in the SPGPE) with experimental observations of the formation of a persistent current. Our results provide an \emph{ab initio} quantitative experimental test of the SPGPE, and indicate the need for both damping and noise terms to give a quantitative account of vortex motion in a high-temperature Bose-Einstein condensate.

\section{c-field theory}
\subsection{Background}
As highly controllable degenerate matter wave systems, dilute Bose gases offer a unique window into the realm of many body quantum mechanics~\cite{Bloch:2008gl}. Yet developing a quantitative non-equilibrium description of high-temperature Bose gases poses a major theoretical challenge~\cite{Proukakis:2008eo,Blakie:2008is}, particularly near the critical temperature where the breakdown of mean-field theory renders two-fluid theories~\cite{Zaremba1999} inoperative. Exact methods offer insight for small systems~\cite{Dagnino09a}, while positive-P \cite{Drummond:2004iw} and Monte Carlo \cite{Kashurnikov:2001et} methods have been applied to determining equilibrium properties. For temperatures near the critical point of evaporative cooling, experimental tests of theory have been confined to collective modes~\cite{Morgan:2003bv,Jackson:2002bu}, equilibrium critical fluctuations~\cite{Davis:2006ic}, and spontaneous vortex formation \cite{Weiler:2008eu}.

The SPGPE used in this work is a grand-canonical c-field method formulated from a microscopic derivation of reservoir interactions in the Wigner phase-space representation \cite{SGPEI,Gardiner:2003bk}, and is valid right through the phase transition~\cite{Bradley:2008gq,Weiler:2008eu,Blakie:2008is}. The equation of motion resembles the projected Gross-Pitaevskii equation (PGPE)~\cite{Davis2001b,Blakie05a,Wright:2008ha,Wright:2009eh}, but also contains damping and noise terms arising from the reservoir interaction, and evolves both the condensed and non-condensed fractions of the Bose field lying below a specified energy cutoff. The cutoff is a central formal and technical aspect of the theory, allowing its consistent extension beyond one spatial dimension~\cite{Gardiner:2003bk} (note also the non-projected SGPE of Refs.~\cite{Stoof1999,Cockburn10a,Cockburn:2012gc,Cockburn09a,Cockburn:2011kw}). The SGPE has also been used to treat systems where a quantitative description of a reservoir with definite atom number is not required~\cite{Damski10a,Das:2012ki,Su:2013dh}.

In this work we use the \emph{simple-growth} SPGPE~\cite{Bradley:2008gq,Weiler:2008eu,Rooney:2010dp,Rooney:2011fm,Rooney:2012gb,Garrett:2013gk}, a treatment that provides a description of a reservoir with definite temperature and atom number, but neglects reservoir interactions involving number-conserving scattering between the classical field and thermal reservoir atoms \cite{Rooney:2012gb}.  These processes induce energy damping and diffusion, and are known to be weak in quasi-equilibrium situations \cite{Blakie:2008is}. In a simple-growth SPGPE study of vortices occurring spontaneously during the phase-transition~\cite{Weiler:2008eu}, a single fitted parameter (the reservoir coupling strength) was used to give vortex formation data in close agreement with the experiment. However, in near-equilibrium situations the reservoir interaction parameters can be determined {\em a-priori} \cite{Bradley:2008gq,Rooney:2010dp,Cockburn10a} allowing the SPGPE to perform quantitatively accurate calculations of dissipative Bose-gas dynamics at high-temperature, provided the thermal reservoir is not significantly disturbed. 

\begin{figure}[t!]
\begin{center}
\includegraphics[width=\columnwidth]{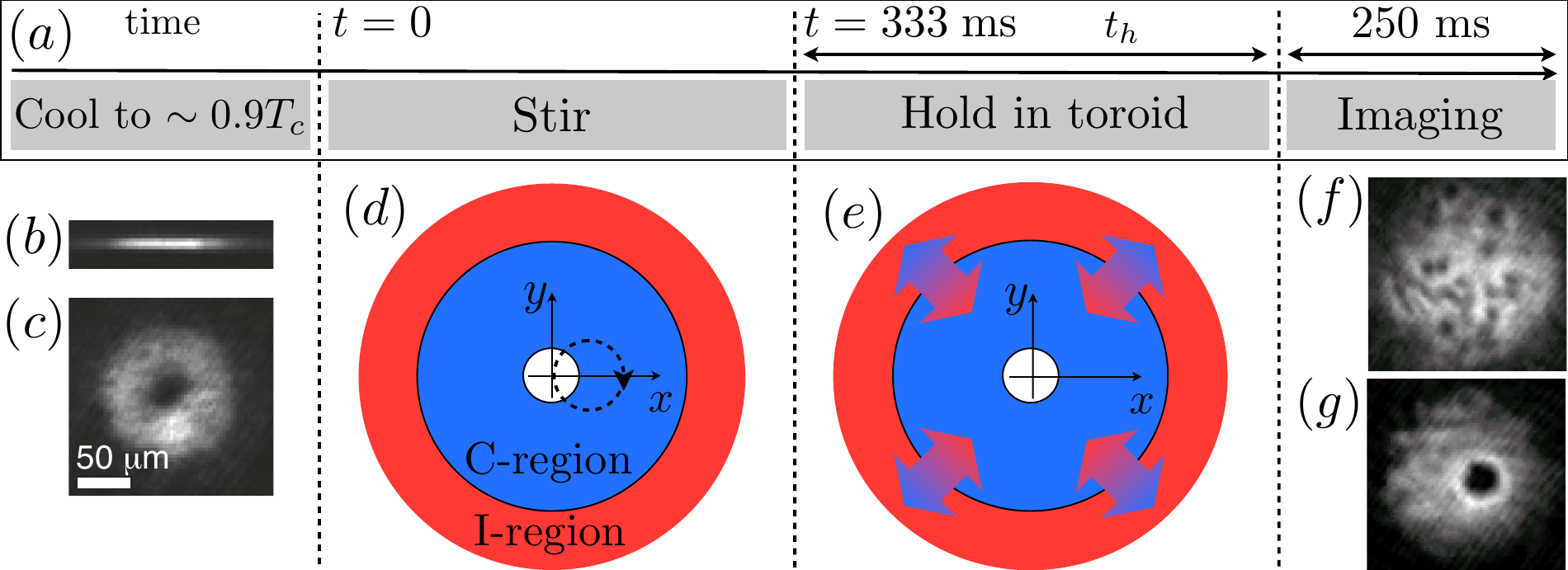}
\caption{(Color online) Schematic of the experimental sequence and simulation parameters~\cite{Neely:2012vj}. $(a)$ The initial state consists of a trapped Bose gas at $T\sim 0.9T_C$ in a cylindrically-symmetric harmonic trap augmented with an optically induced Gaussian obstacle potential. The potential is constant in the $z$ direction and coincident with the symmetry axis of the harmonic trap. In-situ absorption images of the experimental atom density are shown in the transverse $(b)$ and axial $(c)$ directions. $(d)$ The center of the harmonic trap is induced to complete a single revolution around the $z$-axis. The obstacle beam executes one circular orbit centered at $(\bar{x}(t),\bar{y}(t))=r_0(1-\cos{\kappa t},\sin{\kappa t})$, with $r_0=2.875\mu{\rm m}$, $\kappa=6\pi\;{\rm s}^{-1}$. This is modeled in the frame of the obstacle beam, with an incoherent ($\rI$-) region coupled to a coherent ($\rC$-) region described with the SPGPE (shown schematically, see text). $(e)$ The dissipation has significant effect during the experimental hold (time $t_h$ after the stir). Absorption images after significant hold times show either $(f)$ many vortex cores  after an additional radial expansion stage, or $(g)$ a large density minimum corresponding to a persistent current (without radial expansion). }
\label{scheme}
\end{center}
\end{figure}

\subsection{SPGPE theory}

In our c-field description the truncated-Winger field is expanded on a basis of harmonic oscillator eigenstates.  The modes of the system are divided into two distinct regions: the \emph{coherent region} (\rC) consisting of modes with energy less than a specified cutoff ($\ecut$), and the \emph{incoherent region} (\rI) which contains the remaining high energy, quasi-equilibrium states.  The \rI-region acts as a thermal reservoir for the \rC-region, and is assumed to be in thermal equilibrium at a temperature $T$ and chemical potential $\mu$, described by a semiclassical Bose-Einstein distribution.  The \rC-region is treated using the truncated Wigner method, where accounting for interactions between the \rC- and \rI-regions leads to a stochastic differential equation for the \rC-region dynamics.  Individual trajectories evolve according to the stochastic differential equation of motion~\cite{Bradley:2008gq} 
\begin{eqnarray}
\hbar d\psi(\x,t)&=&\PC\Big\{(i+\gamma)(\mu-L)\psi(\x,t) dt +\hbar dW(\x,t)\Big\},
\label{SGPEsimp}
\end{eqnarray}
where the projection operator $\PC$ implements the energy cutoff in the basis of harmonic oscillator modes, and the complex Gaussian noise satisfies $\langle dW(\x,t)dW(\x^\prime,t)\rangle =0$, $\langle dW^*(\x,t)dW(\x^\prime,t)\rangle  =(2\gamma k_BT/\hbar)\delta(\x,\x^\prime)dt,$  where $\delta(\x,\x^\prime)=\sum_{n\in \rC}\phi_n\of\x \phi_n^*\of{\x^\prime}$ is the delta-function for the $\rC$-region. The operator $L$ generates the Hamiltonian evolution for the $\rC$-region $L\psi \equiv \left(H_{\rm sp}+g|\psi|^2\right)\psi$, where the single-particle Hamiltonian is $H_{\rm sp}=-\hbar^2\nabla^2/2m+V(\x,t)$, $g = 4\pi \hbar^2 a/m$ characterizes the strength of the atomic interaction, and $a$ is the s-wave scattering length. The terms involving $\gamma$ in Eq.~(\ref{SGPEsimp}) account for the growth of the \rC-region due to S-wave scattering of two \rI-region atoms, and the corresponding time reversed process.

Experiments are most commonly described by the system temperature and total atom number $N_T$.  In the SPGPE implementation, the choice of $\mu(T,N_T)$ controls the total atom number $N_T$, while the choice of $\ecut (T,N_T)$ dictates the occupation at the cutoff which must be of order unity for the classical field description of the \rC-region to be valid \cite{Blakie:2008is}. A Hartree-Fock method can be used to accurately estimate these SPGPE parameters for harmonically trapped systems close to equilibrium \cite{Rooney:2010dp}.  The growth rate $\gamma$ can be calculated when the incoherent region can be described by an ideal semiclassical Bose-Einstein distribution, an approximation which is reasonable for near equilibrium situations.  The rate is given by \cite{Bradley:2008gq}
\begin{eqnarray}\label{gamdef} \gamma (T,\mu,\ecut) &=&\gamma_0\sum_{k=1}^\infty\; \frac{e^{\beta\mu(k+1)}}{e^{2\beta\ecut k}}\Phi\left[\frac{e^{\beta\mu}}{e^{\beta\ecut}},1,k\right]^2,\label{eqn:dampingrate}
\end{eqnarray}
where $\Phi[u,v,w]$ is the Lerch transcendent and the dimensionless rate constant is $\gamma_0 = 8a^2/\lambda_{dB}^2$ for de~Broglie wavelength $\lambda_{dB}=\sqrt{2\pi\hbar^2/mk_BT}$.  Given this explicit form for $\gamma$, all SPGPE parameters are determined from experimental data prior to simulation, giving a first-principles treatment of damping with no fitted parameters.  

\subsection{Numerical procedure and analysis}\label{numerics}

\begin{figure*}[tpb]
\begin{center}
\includegraphics[width=0.9\textwidth]{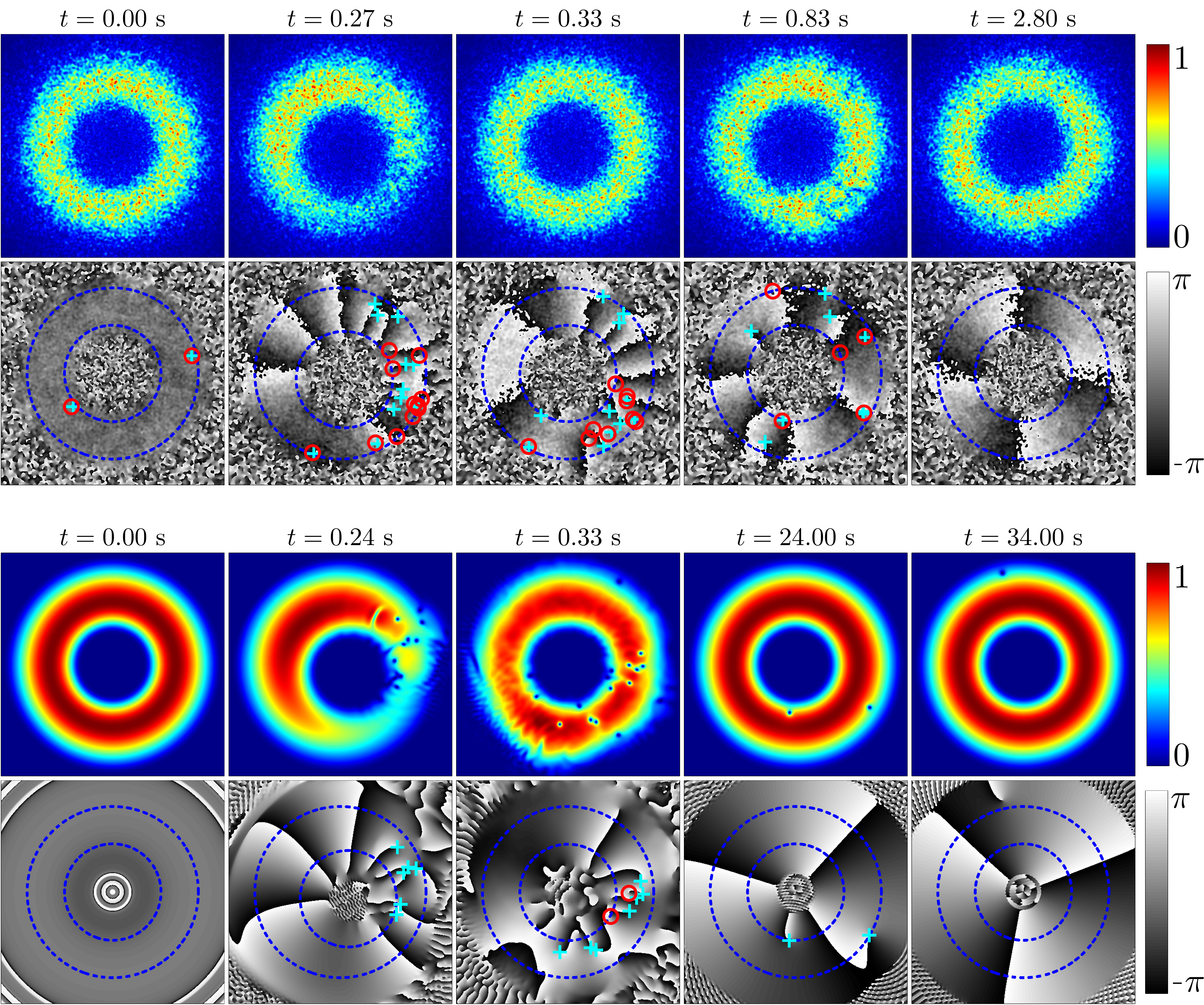}
\caption{(Color online) Column densities and phase slices (through the $z = 0$ plane) showing the c-field dynamics of modeling the persistent current formation experiment, with parameter set (a), $V_0=58\hbar\bar{\omega}$. The blue dashed circles show the boundary of the detection region for vortices that are labelled in the phase profile as positive (cyan plus) or negative (red circle) based on their circulation. Each image is $94\times94~\mu{\rm m}$.  Rows 1 and 2 show SPGPE evolution for a single trajectory~\cite{Note2}. Rows 3 and 4 show the dGPE evolution. }
\label{figsequence}
\end{center}
\end{figure*}
We describe the toroidal system of Ref.~\cite{Neely:2012vj} using a harmonic-Gaussian external potential $V(\x,t) = V_{HO} (\x) + V_G (\x,t)$, where $V_{HO}(\x) = \frac{m}{2} \left[\omega_r^2(x^2+y^2)+\omega_z^2z^2\right]$ is the harmonic oscillator potential, and the time dependent Gaussian potential is given by $V_G(\x,t)=V_0\exp\left[{-[(x-\bar{x}(t))^2+(y-\bar{y}(t))^2] / \sigma_0^2}\right]$. We use the parameters measured in the experiment: $(\omega_r,\omega_z) = 2\pi\times (8,90)$~Hz,  $\sigma_0 = 23/ \sqrt{2}~\mu m $, where $\bar{\omega}  = (\omega_r^2 \omega_z)^{1/3}$. We expect that the potential height $V_0$ may have a significant influence on the measurements results, since the height of the obstacle changes the number of vortices that may be pinned~\cite{TWNThesis}. To account for this variation we run two sets of SPGPE trajectories. Changing $V_0$ requires a change in $\mu$ to preserve $N_T$, and in $\ecut$ to maintain the same population at the cutoff energy ($n_{\rm cut}\approx 1$, see Section \ref{disc}). We choose two values of $V_0$ at the upper and lower values of the experimental uncertainty in the measured value respectively. We then  create an initial equilibrium state by evolving Eq.~\eref{SGPEsimp} with the Gaussian potential at $(\bar{x},\bar{y}) = (0,0)$, and verify its properties. Our parameter sets are:
\par
(a) $V_0 = 58 \hbar \bar{\omega}$, $\mu=34\hbar\bar{\omega}$, $\ecut=83\hbar\bar{\omega}$,\par
(b) $V_0 = 67 \hbar \bar{\omega}$, $\mu=35\hbar\bar{\omega}$, $\ecut=84\hbar\bar{\omega}$.\par
These self-consistently determined parameters allow us to sample the equilibrium ensemble of the SPGPE, for a total of $N_T = 2.6\times10^6$ $^{87}{\rm Rb}$ atoms in the toroidal trapping potential at temperature $T=98{\rm nK}$~\footnote{The details of a self-consistent method of accounting for the above-cutoff atom number are given in Ref.~\cite{Rooney:2010dp}.}, matching the experimental values for atom number and temperature. 
\par
To model the dynamics, the Gaussian obstacle is shifted to $(\bar{x}(t),\bar{y}(t)) = r_0(1-\cos(\kappa t),\sin(\kappa t))$, moving it in a circle of radius $r_0 = 2.875~\mu {\rm{m}}$, about the point $(x,y) = (r_0,0)$, with angular frequency $\kappa = 2 \pi / (333~{\rm{ms}}) = 6\pi~{\rm{s}}^{-1}$ (see \fref{scheme}), as is done in the experiment. After one circular orbit, the potential is held at the trap center. Using Eq.~\eref{eqn:dampingrate}, both parameter sets (a) and (b) give $\gamma = 8 \times 10^{-4}$, setting the dissipation rate during the dynamics. For each parameter set we propagate 16 trajectories of the SPGPE. Our implementation of the damped GPE is identical to that of the SPGPE, except that the noise is set to zero in Eq.~(\ref{SGPEsimp}). 

\begin{figure*}[t!]
\begin{center}
\includegraphics[width= 0.9\textwidth]{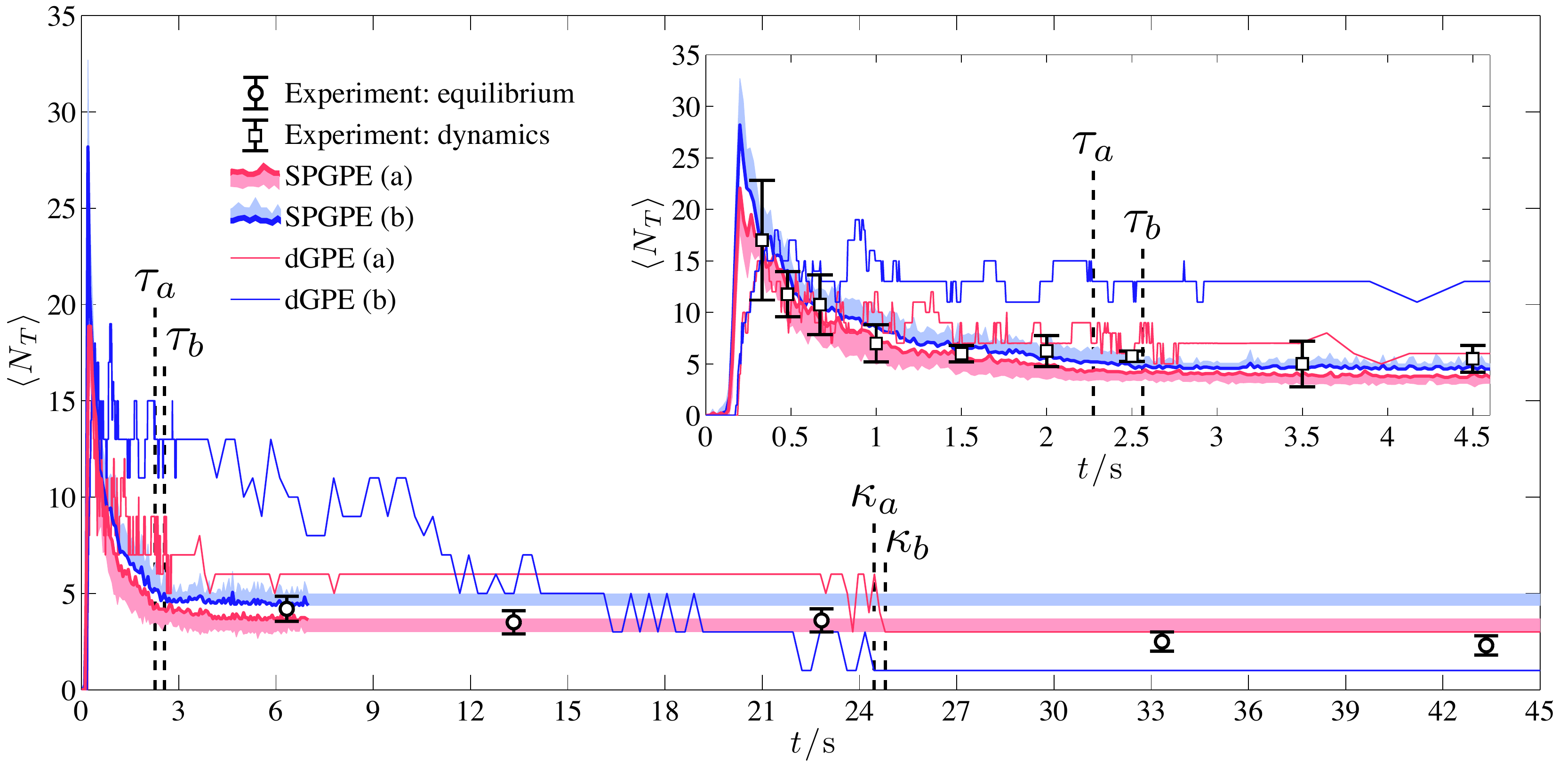}
\caption{(Color online) Comparison of the total vorticity $\langle N_T \rangle$ [\eeref{vortexsum}] between SPGPE theory and experiment. In the main figure the SPGPE ensemble average is shown for $V_0 = 58 \hbar \bar{\omega}$ (SPGPE (a), red thick curve), and $V_0 = 67 \hbar \bar{\omega}$ (SPGPE (b), blue thick curve), and compared with experimental data (equilibrium). In both cases a stable persistent current is formed after $t \sim 3$~s.  The dGPE simulations corresponding to $V_0 = 58 \hbar \bar{\omega}$ (dGPE (a), red thin line), and $V_0 = 67 \hbar \bar{\omega}$ (dGPE (b), blue thin line) develop a stable persistent current after $t \sim 25$~s.   The inset shows the short-time dynamics. The average times at which there are no free vortices after the stir, are shown by the dashed vertical lines at $t = \tau$ for the SPGPE, and $t = \kappa$ for the dGPE, where the subscript denotes the parameter set $(a)$ or $(b)$. The shading for each SPGPE curve shows one standard deviation, where the lower bound for parameter set (a), and upper bound for parameter set (b) are shown for clarity.}
\label{figvortexnumber}
\end{center}
\end{figure*}

We quantitatively compare the SPGPE and dGPE by calculating the number of vortices at a given time in the numerical simulations, and comparing this with the number found in the experiment. Experimentally, the vortices were counted from time-of-flight images after ramping down the Gaussian beam over 250 ms. The experimental data are analyzed differently for the time interval immediately after the stir ($t\lesssim 4.5$s), and for later times where the average number of vortices reaches a quasi-equilibrium state; we refer to the latter interval as \emph{equilibrium}, but it should be noted that even at the longest observation times of the experiment there remains a small rate of free vortex observation, and a slow decline of the mean observed winding number, and hence true equilibrium does not occur. However, the mean total number of vortices reaches equilibrium in our simulations, and quasi-equilibrium in the experiment, after $\sim 4.5$s. After this equilibrium is achieved, the winding number of the persistent current is experimentally determined by introducing an extra 3 s hold after ramping down the obstacle. This allows multiply charged vortices that are no longer stabilized by the obstacle beam to break up into individual vortices of unit circulation \cite{TWNThesis,Murray:2013wq}, that are readily observed. For the short-time data ($t\lesssim 4.5$ s), we use a different technique. We again count free vortices, but do so immediately after ramping down the obstacle.  Without the hold time, multiple vortices simultaneously pinned to the obstacle do not decay into individual vortex cores. Instead, the density minimum at the center of the flow that formed the multiply charged core acquires a new area that depends on the winding number. As recently shown, the area of the density minimum at the core is proportional to the winding number~\cite{Murray:2013wq};  we use this relationship to infer the pinned winding number prior to obstacle ramp-down.
The observed vortex number thus includes free vortices and any vortices pinned to the Gaussian beam prior to the imaging sequence. 
\par
In our simulations we extract the total vorticity to be compared with the experimental data via 
\eqn{N_T =N_f + \mathcal{W}_B\equiv  N_f +  \left| \frac{m}{\hbar} \oint_{B} {\bf v}(\x) \cdot d{\bf l} \right|, \label{vortexsum} }
where $N_f = N_++N_-$ is the number free of vortices (of both positive and negative circulation) in the region of detection, and the $B$ is the inner boundary of the toroidal atomic density, so that the winding number from pinned vortices is $\mathcal{W}_B$.  In equilibrium, $N_T=\mathcal{W}_B$ corresponds to the size of the stable persistent current. To avoid counting thermal fluctuations as vortices, we limit our region of vortex detection to radii of significant atomic density (After the stir, this is the region $22.9 ~\mu {\rm m} < r  < 34.4 ~\mu {\rm m}$, as indicated in \fref{figsequence} by blue dashed circles), consistent with the limitations of experimental vortex detection. 

During the post-stir hold period [Fig.~\ref{scheme} (e)], there is an additional cooling stage of the experiment, however this has a negligible effect on the vortex dynamics of the simulations. A discussion of this and other technical features of our simulations and experiment is given in Section \ref{disc}.

\section{Results}
\fref{figsequence} shows a comparison of individual trajectories of the SPGPE and dGPE for parameter set (a). The dynamics of both methods are qualitatively similar.  Multiple vortices are nucleated during the stirring procedure, and then decay through a range of processes: decay to the exterior condensate boundary, internal vortex-antivortex annihilation, or via pinning at the central potential~\footnote{See Supplemental Material at [URL will be inserted by publisher] for Movie S1 showing a single SPGPE trajectory.}.  Finally the system evolves into a stable persistent current of winding number $N_T = \mathcal{W}_B$, with no free vortices in the bulk fluid.  The  timescale of the decay of free vortices in SPGPE and dGPE differ by an order of magnitude. The SPGPE evolution generates a stable ${\cal W}_B=4$ persistent current after 2.8~s. In contrast, the dGPE requires over 25~s to evolve into a stable persistent current with ${\cal W}_B=3$. 

In \fref{figvortexnumber} we plot $\langle N_T \rangle$ for the SPGPE ensemble average, and for the dGPE simulations. We first show the long-time data in the main plot of \fref{figvortexnumber}, which includes data for the experiment near equilibrium where a stable persistent current has formed.  The timescale of equilibration, i.e. the average time at which $N_f = 0$, is shown for the SPGPE $(\tau)$ and dGPE $(\kappa)$.   Both SPGPE calculations agree well with the first experimental data point at $t = 6.33$~s, lying within the experimental uncertainty.  At this stage $\langle N_T \rangle$ has reached a stable value for the SPGPE simulations and all free vortices have left the condensate, leaving a persistent current.  The SPGPE calculations agree well with the experiment for $6.33 {\rm s}<t<\kappa$. In contrast, free vortices exist in the dGPE simulations for the first 25~s of evolution with $\kappa/\tau \approx 11$ for both parameter sets. Eventually the dGPE evolves into equilibrium with ${\cal W}_B=3$ for (a), and ${\cal W}_B=1$ for (b), differing from the experimental observations, and indicating a high sensitivity to the precise value of $V_0$ used. Note that the long time decay of the experimental value of $\langle N_T \rangle$ is possibly due to a slow drift in the magnetic trap center, rather than the decay of free vortices~\cite{Neely:2012vj,TWNThesis}.

The short-time dynamics (during the stir) are shown in the inset of \fref{figvortexnumber}, where we compare the numerical results with experimental data.  The early time dynamics show a rapid initial rise in $\langle N_T \rangle$ as angular momentum is injected by the stir, with more vortices nucleated in the SPGPE simulations.  The peak number of vortices occurs at $t = 0.21$~s in both SPGPE ensemble averages, after which there is a dramatic drop in $N_f$.  In comparison, the dGPE vortex number peaks at $t = 0.33$~s, followed by a very slow decline.  In general, the SPGPE results agree well with experiment for both parameter sets (a) and (b), while the dGPE shows a slower decay for $\langle N_T \rangle$, and a strong dependence on $V_0$. 

While the dGPE describes the dynamics qualitatively, the the timescale of stable persistent current formation is a factor of 10 slower than observed in SPGPE, and at least a factor of 3  slower than observed in the experiment.    These quantitative differences are our main result: the discrepancies in the dGPE signify a breakdown of the dGPE validity, which in combination with the accuracy of the SPGPE show that the noise is necessary to \emph{quantitatively} reproduce the available experimental data, in both the non-equilibrium and equilibrium stages of the evolution.

In \fref{figvortexnumber} we see that the main effect of varying $V_0$ is to change the size of persistent current formed.  Within the dGPE a larger barrier height leads to a smaller persistent current, ${\cal W}_B=1$, much smaller than the experimentally observed value of $3\leq {\cal W}_B\leq 5$. In contrast to the dGPE, in SPGPE a larger barrier leads to a \emph{larger} persistent current; this suggests that fluctuations enhance vortex mobility near the barrier. 

\section{Discussion}\label{disc}
In this section we discuss the consistency of our simulations, and aspects of vortex imaging, the cooling sequence, and the role of experimental uncertainties. 
\subsection{SPGPE Simulations}
\subsubsection{Consistency of simulations}
Our numerical method is a spectral Galerkin method based on Gauss-Hermite quadrature \cite{Blakie08b,Blakie:2008is}.  Due to large particle number, the trap oblateness, and the need for a high energy cutoff, we require $\sim 10^5$ modes in the $\rC$-region, making each trajectory numerically challenging~\footnote{Our numerical method implements the energy cutoff very accurately, but with a numerical cost that scales rapidly with increasing cutoff energy~\cite{Blakie08b}. For the large system described in this work, each trajectory of the SPGPE takes $\sim 50$ days of wall clock time to propagate for $7$ s of system evolution, on the University of Otago Vulcan computing cluster with 2.66 GHz CPUs.}. However our method is advantageous since our choice of basis allows us to precisely implement a consistent energy cutoff for this system in the single-particle basis, since at sufficiently high energies the many-body Hamiltonian is diagonalized by this basis.  

Since the SPGPE is a formally projected theory, the cutoff independence should be verified before make quantitative predictions. For our system, the chosen cutoff energy $\ecut$ gives an average occupation at the cutoff of $n_{\rm cut} \approx 1$. Cutoff independence was checked by performing simulations with $\ecut$ lowered by 12\%. This resulted in a similar mean cutoff occupation of $n_{\rm cut} \approx 1.1$, and no discernible differences in the dissipative evolution of SPGPE simulations.  Note also that the Gaussian potential is well represented in the basis of single-particle states defining the \rC-rgeion due to a separation of energy and length scales: $V_0 \ll \ecut$, and $\sigma_0 \ll R_{\rm cut}= \sqrt{2 \ecut / m\omega_r^2} = 73~\mu{\rm m}$. Thus our basis gives a complete representation of the $\rC$-region field in the combined trap.

\subsubsection{Limitations}
There are two simplifications of our SPGPE treatment that require further discussion. Firstly, we have neglected the dynamics of the atoms in the \rI-region. In general this can be a significant effect \cite{Jackson:2002bu}, but in the system we consider the thermal fraction is very large ($\sim 70\%$), justifying the approximation that the effect of the condensate on the thermal cloud is negligible. From a technical standpoint, a theoretical framework that encompasses SPGPE dynamics for the \rC-region and includes dynamics of the \rI-region is yet to be developed. Secondly, we have neglected the so-called \emph{scattering} terms~\cite{Rooney:2012gb}, involving the number-conserving exchange of energy between \rC- and \rI-regions. Such terms are significant when the \rC- and \rI-regions are far from mutual equilibrium, so it is possible that these terms have a significant effect during the stir. However we see that the simple growth model accurately reproduces the available experimental data during the short time evolution.  The accuracy of our SPGPE simulations is because the system rapidly evolves into quasi-equilibrium [see Fig.~\ref{figsequence}-\ref{figvortexnumber}], ensuring the validity of the simple growth model.

\subsection{Experimental Features and Analysis}\label{sec:exp}

\subsubsection{Vortex imaging}
Vortex imaging involves ramping down the obstacle potential, introducing the possibility that vortices initially pinned to the barrier could decay prior to imaging. Simulations of this ramp down show the vortex number is preserved, as can be expected on physical grounds since the vortex decay time is much longer than the ramp time. The time-of-flight imaging also does not  change the vortex number, as it occurs after the trap potential is snapped off, causing ballistic expansion of the cloud and rapid extinguishing of any further mean-field dynamics. Preservation of vortex number during imaging has also been verified in experiments with identical expansion parameters performed in Ref.~\cite{TWNThesis,Neely:2010gl} where vortex numbers consistent with known forcing were observed.  

\subsubsection{Extra cooling}
As discussed in Section \ref{numerics}, there is an additional cooling stage at the end of the experimental sequence that we do not present data for here. 
After $t = 1.5$ s the system is cooled to $T \sim0.6T_c$, with $N_T = 1\times10^6$~\cite{Neely:2012vj}. We have simulated this cooling by instantaneously changing reservoir parameters at $t = 1.5$~s to $(T, \mu, \ecut, \gamma) = (47 {\rm nK},33 \hbar \bar{\omega}, 83 \hbar \bar{\omega}, 0.8\times10^{-4})$, to give a reservoir atom number and temperature consistent with the end state of the experimental sequence.  Modeling this process revealed no significant modification to the SPGPE value for $\langle N_T\rangle$, consistent with a quasi-equilibrium state associated with a stable winding number being reached before the cooling process begins. 

\subsubsection{Experimental uncertainties}
In addition to the main parameter we have considered ($V_0$), there is further experimental uncertainty in other parameters including atom number, temperature, stirring velocity, beam width, and beam position.  Numerical simulations with variations in these parameters would also lead to variations in the final persistent current size.  However we expect that the most significant change will be due to including thermal driving noise in the SPGPE, which significantly increases vortex decay rates~\cite{Rooney:2010dp}. Thus the persistent current formation time in SPGPE can be expected to be much faster than in dGPE, irrespective of these various experimental uncertainties.    

\section{Conclusions}
In this work we have modeled the persistent current experiment of Neely \emph{et al.} \cite{Neely:2012vj} using a grand-canonical c-field theory of reservoir interactions. We have performed an \emph{ab initio} quantitative test of the SPGPE with no fitted parameters. The SPGPE theory quantitatively reproduces the dynamics observed in the experiment at short times, and accurately predicts the experimentally observed persistent current formation time \emph{and} winding number.  While the damped GPE is qualitatively correct it is not quantitatively informative, as could be expected in the high-temperature regime of the experiment $(T \approx 0.9 T_c)$. In general, both damping and noise are required to give a quantitative description of dissipation in open quantum systems, and our results demonstrate the central importance of thermal noise in high-temperature Bose-gas dynamics, with particular emphasis on the motion of quantized vortices. Our approach provides a general quantitative framework for modelling the high-temperature dynamics of trapped Bose-Einstein condensates.

\section*{Acknowledgments}
We thank Crispin Gardiner for a critical reading of this manuscript.
SJR was supported by The University of Otago, and thanks Victoria University where this work was finished for their hospitality. ASB was was supported by a Rutherford Discovery Fellowship administered by the Royal Society of New Zealand, and the Marsden Fund.  The experimental work was supported by the US National Science Foundation.


\end{document}